\documentclass[12pt]{article}
\usepackage{amssymb}
\usepackage{amsmath}
\usepackage{hyperref}
\begin{document}
\title{\bf  AdS Black Brane Solution Surrounded by Quintessence in Massive Gravity and KSS Bound}

\author{Mehdi Sadeghi\thanks{Corresponding author: Email: mehdi.sadeghi@abru.ac.ir}  \hspace{2mm} \\
		 {\small {\em  Department of Physics, School of Sciences,}}\\
		 {\small {\em  Ayatollah Boroujerdi University, Boroujerd, Iran}}\\
    }

\date{\today}
\maketitle

\abstract{In this paper, the Einstein AdS black brane solution in the presence of quintessence in context of massive gravity is introduced. The ratio of shear viscosity to entropy density for this solution violates the KSS bound by applying the Dirichlet boundary and regularity on the horizon conditions. Our result shows that this value is independent of quintessence in arbitrary dimensions.}\\



\section{Introduction} \label{intro}
 
 Astronomical observations shows that $70 \%$ of the universe is unknown  which is called dark universe. Thus, Physicists are trying to modify general theory of relativity to include dark matter and dark energy. One of these modifications is massive gravity \cite{deRham:2010kj}, to explain dark univers. Astronomical observations have shown that the universe has an accelerating expansion \cite{Perlmutter:1998np,Riess:1998cb}, which might be due to dark energy acting as an anti-gravity. Quintessence is a possible origin of this phenomena with the state equation $p =\omega_{q} \rho_{q}$ \cite{Caldwell:1997ii} where $p$ and  $\rho_{q}$ are the negative pressure and the energy density  respectively\cite{Caldwell:1997ii}. The first general solution of spherically symmetric static Einstein equations in four dimensions in the presence of  quintessence  was introduced by Kiselev \cite{Kiselev:2002dx}. The extension of this solution to d-dimensions was constructed by \cite{Chen:2008ra} on the brane. Other extensions of Kiselev solution were studied in \cite{Azreg-Ainou:2014lua,Ghosh:2016ddh,Ghosh:2017cuq,Chabab:2020ejk}.

AdS/CFT duality relates two different theories that  are physically and mathematically different. This duality    states that super Yang-Mills $\mathcal{N}=4$ with gauge group $SU(N)$ corresponds to string theory $IIB$ in $AdS_5\times S^5$, was introduced by Maldacena in 1997\cite{Maldacena97}. Gravity is located in the bulk of $AdS_5$ space-time and field theory is located on the boundary of $AdS_5$. Therefore, gravity in $(n+1)$-dimensions corresponds to field theory in $n$-dimensions.  This duality arises from  duality between closed and open strings. Perturbative method is not applicable to strongly coupled field theories. Therefore, this duality is a valuable tool to study the near-equilibrium property of strongly coupled systems.\\
 AdS/CFT duality is known as fluid-gravity duality \cite{Bhattacharyya07,Bhattacharya2011} in the long-wavelenght limit. In this regime, field theory is effectively described by hydrodynamics. Quark-Gluon Plasma \cite{Shuryak:2008eq} and cold atomic gases \cite{Cao:2010wa} have hydro-dynamical behavior and it has been  observed experimentally. \\
The result of $\frac{\eta}{s}$ for QGP from fluid-gravity duality is in qualitative agreement with experimental data. This conjecture ,$\frac{\eta }{s} \ge \frac{1}{4\pi}$, is preserved for all relativistic quantum field theories at finite temperature with gravitational dual and this value get a lower bound, $\frac{1}{4\pi}$,  for Einstein-Hilbert gravity which is known as Kovtun-Son-Starinets bound \cite{Policastro2001,Policastro2002,Son2007,Kovtun2004,Kovtun2012}. This conjecture is violated for higher derivative and massive gravity theories\cite{Ref23,Ref24,Sadeghi:2015vaa,Parvizi:2017boc,Sadeghi:2018vrf}. Shear viscosity is one of the transport coefficients appeared at first order perturbation of dissipative fluid.
 \begin{align}
 & T^{\mu \nu } =(\epsilon +p)u^{\mu } u^{\nu } +pg^{\mu \nu } -\sigma ^{\mu \nu },\\
 &\sigma ^{\mu \nu } = {P^{\mu \alpha } P^{\nu \beta } } [\eta(\nabla _{\alpha } u_{\beta } +\nabla _{\beta } u_{\alpha })+ (\zeta-\frac{2}{3}\eta) g_{\alpha \beta } \nabla .u].
 \end{align}
  where $\eta$, $\zeta $, $\sigma ^{\mu \nu }$ and $P^{\mu \nu }$ are shear viscosity, bulk viscosity, shear tensor and projection operator, respectively \cite{Kovtun2012}.\\
  Green-Kubo formula can be derived by linear response theory as follows, 
  \begin{equation}\label{Kubo}
  \eta =\mathop{\lim }\limits_{\omega \to 0} \frac{1}{2\omega } \int dt\,  d\vec{x}\, e^{\imath\omega t} \left\langle [T_{y}^{x} (x),T_{y}^{x} (0)]\right\rangle =-\mathop{\lim }\limits_{\omega \, \to \, 0} \frac{1}{\omega } \Im G_{y\, \, y}^{x\, \, x} (\omega ,\vec{0}).
  \end{equation}
  The shear viscosity $\eta$ is straightforwardly obtained from a Kubo formula and the entropy density $s$ from Bekenstein-Hawking formula. Kubo formula for massive gravity was derived in \cite{Burikham:2016roo}.\\
\indent where $\eta$, $\zeta $, $\sigma ^{\mu \nu }$ and $P^{\mu \nu }$ are shear viscosity, bulk viscosity, shear tensor and projection operator, respectively \cite{Kovtun2012}.\\
In this paper we consider massive gravity with quintessence in d-dimensions and introduce the black brane solution. Finally, we study the effect of bulk dimension, in the presence of quintessence, on the value of $\frac{\eta}{s}$ and put some comments on the fluid  dual to this gravity model.\\
The goal of this paper is to present a model of the universe
in which the dark energy component is described by both a quintessence field and a negative cosmological
constant. The quintessence field accounts for the present
stage, accelerating expansion, of the universe. Meanwhile,
the inclusion of a negative cosmological constant warrants
that the present stage of accelerating expansion will be, eventually,
followed by a period of collapse into a final cosmological
singularity (AdS universe).
 
\section{The Einstein AdS Black Brane Solution Surrounded by Quintessence in Massive Gravity }
 \label{sec2}
The action for this model is written as follows,
\begin{align}\label{action}
S&=\frac{1}{2}\int{d^dx\sqrt{-g}\Bigg[R-2\Lambda+m_g^2\sum_{i=1}^{d-2}{c_{i}\mathcal{U}_i(g,f)}\Bigg]}+S_{\text{quint}},
\end{align}
The first term,$ R $, is the Ricci scalar and it represent the Einstein gravity. The second term ,$\Lambda$, is the cosmological constant expressed
in terms of the curvature radius $l$ of the AdS spacetime background as,
\begin{equation}
\Lambda=\frac{-d_1d_2}{2l^2},\,\, \text{where}\,\,\,\,\, d_i=d-i
\end{equation}
 and it should be negative because we are in AdS space-time. The third term is massive term where $m_g$ is the mass of graviton, $ c_i $'s are coupling parameters, $f$ is a fixed reference metric defined by 	$ f_{\mu \nu} = diag(0,0,c_0^2 h_{ij})$ and $h_{ij}=\frac{1}{l^2}\delta_{ij}$ {\cite{Cai:2014znn} , $ \mathcal{U}_i $ are symmetric polynomials in
 terms of the eigenvalues of the $ d\times d $ matrix $ \mathcal{K}^{\mu}_{\nu}=\sqrt{g^{\mu \alpha}f_{\alpha \nu}} $ given as,
\begin{align}\label{U} 
  & \mathcal{U}_1=[\mathcal{K}],\,\,\,\,\,\,\,\,\mathcal{U}_2=[\mathcal{K}]^2-[\mathcal{K}^2],\,\,\,\,\,\,\,\mathcal{U}_3=[\mathcal{K}]^3-3[\mathcal{K}][\mathcal{K}^2]+2[\mathcal{K}^3]\nonumber\\
  & \mathcal{U}_4=[\mathcal{K}]^4-6[\mathcal{K}^2][\mathcal{K}]^2+8[\mathcal{K}^3][\mathcal{K}]+3[\mathcal{K}^2]^2-6[\mathcal{K}^4]\nonumber \\
 &
 \mathcal{U}_5=[\mathcal{K}]^5-10[\mathcal{K}^2][\mathcal{K}]^3+20[\mathcal{K}^3][\mathcal{K}]^2-20[\mathcal{K}^2][\mathcal{K}]^3+15[\mathcal{K}][\mathcal{K}^2]^2-30[\mathcal{K}][\mathcal{K}]^4+24[\mathcal{K}^5]\nonumber \\
&...
\end{align}
The rectangular brackets denote traces.\\
By calculating of $ \mathcal{U}_i $, they can be written as,
\begin{equation}
\mathcal{U}_i=(\frac{c_{0}}{r})^i\prod_{j=2}^{i+1}d_j,
\end{equation}
in which $\displaystyle{\prod_{x}^{y}...=1}$ if $x>y$.\\
The last term in (\ref{action}) , $S_{\text{quint}}$ , is the action for quintessence matter. A general action
for quintessence in d-dimensional space-time is
\begin{equation}\label{Quin}
S_{\text{quint}}=\int{d^dx\sqrt{-g}\Bigg[-\frac{1}{2}(\vec{\nabla} \phi)^2-V(\phi)\Bigg]}
\end{equation}
The energy-momentum tensor of the quintessence dark energy in an arbitrary dimension can be described by
\cite{Chen:2008ra} ,
\begin{align}
T_{t}\,^{t}&=T_{r}\,^{r}=\rho_q=-\frac{\omega \alpha d_1 d_2 }{4 r^{d_1 (\omega +1)}},\\
T_{x_i}\,^{x_i}&=-\frac{\rho_q}{d_2}(d_1 \omega_q+1),\,\, i=(1,...,d_2)
\end{align} 
We get the following metric as an ansatz for a d-dimensional planar AdS black brane,
 \begin{equation}\label{metric}
 ds^{2} =-f(r)dt^{2} +\frac{dr^{2}}{f(r)} +r^2\sum_{i=2}^d h_{ij}dx^idx^j,
 \end{equation}
 The equations of motion can be derived by variation the action (\ref{action}) with respect to the metric,
  \begin{equation}\label{EoM}
  {G_{\mu \nu }} + \Lambda {g_{\mu \nu }} -{{m}^2}{\chi _{\mu \nu }} = {2T_{\mu \nu }\text{(quint)}}
  \end{equation}
where $G_{\mu \nu }= R_{\mu \nu}-\frac{1}{2}g_{\mu \nu}R$  is the Einstein tensor, $T _{\mu \nu }\text{(quint)}$ is the energy-momentum tensor of quintessence and  ${\chi _{\mu \nu }}$  is massive term,
  \begin{align}\label{chi}
  \mathcal{\chi}_{\mu \nu} =\frac{c_1}{2}\bigg(\mathcal{U}_1 g_{\mu \nu }-\mathcal{K}_{\mu \nu}\bigg)+\frac{c_2}{2}\bigg(\mathcal{U}_2 g_{\mu \nu }-2\mathcal{U}_1\mathcal{K}_{\mu \nu}+2\mathcal{K}^2_{\mu \nu}\bigg)+\frac{c_3}{2}\bigg(\mathcal{U}_3 g_{\mu \nu }-3\mathcal{U}_2 \mathcal{K}_{\mu \nu}\nonumber\\+6\mathcal{U}_1 \mathcal{K}^2_{\mu \nu}-6\mathcal{K}^3_{\mu \nu}\bigg)+\frac{c_4}{2}\bigg(\mathcal{U}_4g_{\mu \nu }-4\mathcal{U}_3 \mathcal{K}_{\mu \nu}+12\mathcal{U}_2 \mathcal{K}^2_{\mu \nu}-24\mathcal{U}_1 \mathcal{K}^3_{\mu \nu}+24\mathcal{K}^4_{\mu \nu}\bigg)\nonumber\\+\frac{c_5}{2}\bigg(\mathcal{U}_5g_{\mu \nu }-5\mathcal{U}_4 \mathcal{K}_{\mu \nu}+20\mathcal{U}_3 \mathcal{K}^2_{\mu \nu}-60\mathcal{U}_2 \mathcal{K}^3_{\mu \nu}+120\mathcal{U}_1\mathcal{K}^4_{\mu \nu}-120\mathcal{K}^5_{\mu \nu}\bigg)+...
  \end{align}
By using this metric ansatz (\ref{metric}), we have:
\begin{align}
R_{tt}&=-\frac{f(r)\bigg(f''(r)r+d_2 f'(r)\bigg)}{2r}\nonumber\\
R_{rr}&=\frac{f''(r)r+d_2 f'(r)}{2f(r)r}\nonumber\\
R&=\frac{f''(r)r^2+2d_2r f'(r)+d_2d_3f(r)}{r^2} \nonumber\\
G_{tt}&=R_{tt}-\frac{1}{2}g_{tt}R=\frac{d_2f(r)\bigg(f'(r)r+d_3 f(r)\bigg)}{2r^2}\nonumber\\
\chi_{tt}&=f(r)m_g^2\bigg(\sum_{i=1}^{d-2}\frac{c_0^ic_i}{2r^{i}} \prod_{j=2}^{i+1} d_j\bigg)\nonumber\\
\end{align}
We will obtain a differential equation for $f(r)$ by using $tt$ component of Eq.(\ref{EoM}) as following,
\begin{equation}\label{eom}
d_2 d_3(k-f(r))-d_2 r f'(r)-2\Lambda r^2+m_g^2\bigg(\sum_{i=1}^{d-2}\frac{c_0^ic_i}{r^{i-2}} \prod_{j=2}^{i+1} d_j\bigg)=-\frac{\omega \alpha d_2 d_1 }{ r^{-2+d_1 (\omega +1)}}   
\end{equation}  
Multiplying the above equation by a factor of $\frac{r^{d_4}}{d_2}$ gives us,
\begin{equation}
d_3r^{d_4}f(r)+ r^{d_3} f'(r)-k d_3r^{d_4} +\frac{2\Lambda r^{d_2}}{d_2} -m_g^2\bigg(\sum_{i=1}^{d-2}\frac{c_0^ic_ir^d}{d_2r^{i+2}} \prod_{j=2}^{i+1} d_j\bigg)=-\frac{\omega \alpha d_1 r^{d_4} }{r^{-2+d_1 (\omega +1)}}
\end{equation}
Thus, we get,
\begin{equation}\label{tt}
 \frac{d}{dr}\bigg(r^{d_3} f(r)\bigg)-k d_3r^{d_4} +\frac{2\Lambda r^{d_2}}{d_2} -m_g^2\bigg(\sum_{i=1}^{d-2}\frac{c_0^ic_ir^d}{d_2r^{i+2}} \prod_{j=2}^{i+1} d_j\bigg)=-\frac{\omega \alpha d_1 r^{d_4} }{r^{-2+d_1 (\omega +1)}},
\end{equation}
$f(r)$ will be obtained by solving Eq.(\ref{tt}) as follows,
\begin{equation}\label{}
 f(r)=k-\frac{b}{r^{d_3}}-\frac{2\Lambda}{d_1 d_2} r^2-\frac{\alpha}{r^{d_1 \omega_q+d_3}}+m_g^2\sum_{i=1}^{d-2} \bigg(\frac{c_0^ic_i}{d_{i+1}d_2r^{i-2}} \prod_{j=2}^{i+1} d_j\bigg).
\end{equation}
Where $k$ and $b$ are constant. $b$ is determined by applying the condition of $f(r)$ on event horizon i.e. $f(r_0)=0$.
\begin{equation} 
b=r_0^{d_3}\bigg[k-\frac{\alpha}{r_0^{d_1 \omega_q+d_3}}-\frac{2\Lambda}{d_1 d_2} r_0^2+m_g^2\sum_{i=1}^{d-2} \bigg(\frac{c_0^ic_i}{d_{i+1}d_2r_0^{i-2}} \prod_{j=2}^{i+1} d_j\bigg)\bigg]. 
\end{equation}
By inserting $m$ in $f(r)$ we have,
 \begin{align}\label{f}
&f(r)=\bigg[k\big(1-(\frac{r_0}{r})^{d_3}\big)-\frac{\alpha}{r^{d_3}}(\frac{1}{{r_0}^{d_1\omega_q}}-\frac{1}{r^{d_1\omega_q}})-\frac{2r_0^2\Lambda}{d_1 d_2}\bigg((\frac{r}{r_0})^2-(\frac{r_0}{r})^{d_3}\bigg)\nonumber\\&+m_g^2\sum_{i=1}^{d-2} \bigg(\frac{c_0^ic_i}{d_{i+1}d_2}\big(\frac{1}{r^{i-2}}-\frac{1}{r_0^{i-2} }(\frac{r_0}{r })^{d_3}\big) \prod_{j=2}^{i+1} d_j\bigg)\bigg].
\end{align}
We are finding the black brane solution, thus, set $k=0$. The emblacking factor is as follows,
 \begin{align}\label{f}
 &f(r)=\bigg[-\frac{\alpha}{r_0^{d_1 \omega_q+d_3}}\big(1-(\frac{r_0}{r})^{d_3}\big)-\frac{2r_0^2\Lambda}{d_1 d_2}\bigg((\frac{r}{r_0})^2-(\frac{r_0}{r})^{d_3}\bigg)\nonumber\\&+m_g^2\sum_{i=1}^{d-2} \bigg(\frac{c_0^ic_i}{d_{i+1}d_2}\big(\frac{1}{r^{i-2}}-\frac{1}{r_0^{i-2} }(\frac{r_0}{r })^{d_3}\big) \prod_{j=2}^{i+1} d_j\bigg)\bigg].
 \end{align}
The entropy density is calculated by using Hawking-Bekenstein formula as,
 \begin{equation}
s=\frac{A}{4G V_{d_2}}=\frac{4 \pi}{V_{d_2}} \int{d^{d_2}x \sqrt{-g}}=\frac{4 \pi}{V_{d_2}} \int{d^{d_2}x \sqrt{-\chi(r_0)}}  =4\pi(\frac{r_0}{l})^{d_2}
\end{equation}
where $V_{d_2}$ is the volume of the constant $t$ and $r$ hyper-surface with radius $r_{0}$ , $\chi(r_0)$ is the determinant of the spatial metric on the horizon and we used $\frac{1}{16\pi G} =1$ so $\frac{1}{4G} =4\pi$.
\section{Calculation of $\frac{\eta}{s}$}
\label{sec3}
In this section, we compute  the shear viscosity in
the boundary field theory by applying fluid-gravity duality \cite{Kovtun2003}. We need to have
$[T{^x\,_y}(x),T{^x\,_y} (0)]$ for computing the shear
viscosity by using fluid-gravity duality. 2-point function of energy-momentum tensor in the boundary field theory is constructed by considering small fluctuations around the black brane
background $g_{\mu \nu}\to g_{\mu \nu}+h_{x_1x_2}$.\\
By following the procedure of \cite{Hartnoll:2016tri} we perturbe the metric as $h_{x_1x_2}=\frac{r^2}{l^2}\phi(r)e^{-\imath\omega t}$ in the equation of motion (\ref{EoM}). The equation of motion
for the fluctuations is the following form,
\begin{equation}\label{mode}
\frac{1}{\sqrt{-g}}\partial_r\bigg(\sqrt{-g}g^{rr}\partial_r{\phi}\bigg)+[g^{tt}\omega^2-m(r)^2]\phi=0
\end{equation}
\begin{equation}
m(r)^2=g^{xx}T_{xx}-\frac{\delta T_{xy}}{\delta g_{x_1x_2}}\nonumber
\end{equation}
Shear viscosity is calculated by Eq.(\ref{Kubo}),\\
\begin{equation}
\eta=\mathop{\lim }\limits_{\omega \, \to \, 0} \frac{1}{\omega } \Im G^{R} _{T^{x_1x_2}T^{x_1x_2}} (\omega ,k=0)=\frac{\sqrt{\chi(r_0)}}{16\pi G_N}\phi_0(r_0)^2=\frac{s}{4\pi} \phi_0(r_0)^2
\end{equation}
 Then, we will have,
\begin{equation}\label{formula}
\frac{\eta}{s}=\frac{1}{4\pi} \phi_0(r_0)^2
\end{equation}
where $\phi_0$ is the solution of Eq. (\ref{mode}) at zero frequency ($\omega=0$).\\
We apply two conditions for $\phi$:  it is regular at horizon $r=r_0$ and  goes  to $\phi=1$ near the boundary as $r \to \infty$.\\
So we perturb the metric (\ref{metric}) according to \cite{Hartnoll:2016tri},
\begin{equation}\label{perturb}
ds^2=-\frac{f_1(r)}{l^2}dt^2 + \frac{l^2}{f_1(r)}dr^2 + \frac{r^2}{l^2}(dx_1^2 + dx_2^2 + 2\phi(r)e^{-\imath\omega t}dx_1dx_2+dx_3^2+...+dx_d^2)
\end{equation}
where
\begin{align}
f_1(r)&=\frac{l^2}{r^{d_3}}\bigg[-\frac{\alpha}{r_0^{d_1 \omega_q}}\big(1-(\frac{r_0}{r})^{d_3}\big)-\frac{2\Lambda}{d_1d_2}(r^{d-1}-r_0^{d-1})+m^2\frac{c_0c_1}{d_2}(r^{d_2}-r_0^{d_2}))\nonumber\\&+m^2 l^2 c_0^2c_2(r^{d_3}-r_0^{d_3})+m^2c_0^3c_3(r^{d_4}-r_0^{d_4})d_3+m^2c_0^4c_4(r^{d_5}-r_0^{d_5})d_3d_4\bigg]=l^2 f(r)
\end{align}
By Plugging Eq.(\ref{perturb}) into the action Eq.(\ref{action}) and keeping up to $\phi^2$ terms \cite{Policastro2001,Policastro2002,Son2007,Kovtun2004,Kovtun2012,Kovtun2003} for $\omega=0$, we get:
\begin{equation}\label{perturbed}
S_2=\frac{-1}{2} \int d^4x \Big(K_1 \phi'^2 -K_2 \phi^2\Big)
\end{equation}
where
\begin{align}
K_1&=\frac{r^{d_2}}{l^{d-2}}\frac{ f_1(r)}{l^{2}}=\frac{r^{d_2}}{l^{d}}f_1(r)\nonumber\\
K_2&= \frac{m^2}{2l^{d_2}}\big(c_0c_1 r^{d_3}+d_4c_0^2c_2r^{d_4}+d_4 d_5 c_0^3 c_3r^{d_5}\big)
\end{align}
The EoM of $\phi$ is derived by variation of $S_2$ with respect to $\phi$,
\begin{equation}\label{EoM2}
(K_1 \phi')' + K_2 \phi=0
\end{equation}
Since Eq.(\ref{EoM2}) doesn't have an exact solution, we solve it perturbatively in terms of  $m^2$ and  $\alpha$. For the leading order we consider $m=\alpha=0$. thus we have,
\begin{equation}\label{EoM-zeromass1}
\Bigg(r \left(r^{d-1}-r_0^{d-1}\right) \phi_0'(r)\Bigg)'=0,\,\,\,d \neq 1
\end{equation}
The solution is founded as,
\begin{align}
\phi_0(r)= C_1+C_2 \left(d \log r-\log \left(r r_0^d-r_0 r^d\right)\right)
\end{align}
$C_2=0$ by applying regularity on the horizon and $C_1=1$ by applying 
$\phi(r \to \infty)=1$ so $\phi_0(r)=1$. In this case, according to Eq. (\ref{formula}),
\begin{equation}\label{ValuM,a=0}
\frac{\eta}{s}= \frac{1}{4\pi}\phi_0(r_0)^2 = \frac{1}{4\pi}
\end{equation}
 Now we consider $m^2$  and $\alpha$ to be a small parameter and try to solve Eq.(\ref{EoM2}). By Putting $\phi=\phi_0+m^2\phi_1(r)+\alpha \phi_2(r)$ where $\phi_0=1$ and expanding EoM in terms of powers of $m^2$ and $\alpha$, we will find,
  \begin{equation}
 \frac{m^2}{2l^{d_2}}\big(c_0c_1 r^{d_3}+d_4c_0^2c_2r^{d_4}+d_4 d_5 c_0^3 c_3r^{d_5}\big)+\left(-\frac{2 \Lambda  m^2 r  \left(r^{d_1}-r_0^{d_1}\right) \phi_1
  	'(r)}{l^{d}d_2 d_1}\right)'=0
   \end{equation} 
  
\begin{align}
\alpha \Bigg(r\left( r^{d-1}- r_0^{d-1}\right) \phi_2''(r)+\left(d  r^{d-1}-r_0^{d-1}\right) \phi_2'(r)\Bigg)=0
\end{align}

 Thus we find the solutions as follows,
  \begin{align}
\phi_1 (r) &\to C_2 + \frac{C_1 \left(d \log (r)-\log \left(r r_0^d-r_0 r^d\right)\right) }{r_0^{-d}d_1}\nonumber\\&-\int^r \frac{c_0 d_2 d_1 r_0 U^{d_4} \left(\frac{c_0 c_2 d_4
   U}{d_3}+\frac{c_1 U^2}{d_2}+c_0^2 c_3 d_5\right)}{4\Lambda \left(r_0^d U-r_0 U^d\right)} \, dU
 \end{align}  
 \begin{align}\label{Phi2}  
\phi_2(r)\to C_3+\frac{C_4 }{r_0^{d}d_1} \left(d \log (r)-\log (r r_0^d-r_0 r^d)\right)
 \end{align} 
 $\phi(r)$ is as follows,  

  \begin{align}\label{Phi-}
\phi(r) =& \phi_0+\alpha C_3+m^2 C_2 +\frac{ \left(\alpha C_4+m^2C_1 \right) \left(d \log r-\log \left(r r_0^d-r_0 r^d\right)\right)}{r_0^d d_1}\nonumber\\&-\frac{c_0
 d_1 m^2 r_0 l^{d-4} \left(\int \frac{r^{d_3} \left(c_0 c_2 d_4 d_2^2+ d_3 c_1 r\right)}{r r_0^d-r_0 r^d} \,
 	dr\right)}{ 2d_3 d_2 \Lambda }
  \end{align}
 where $\Phi_0=\phi_0+\alpha C_3+m^2 C_2$.\\
 
 Near horizon of Eq. (\ref{Phi-}) is as,
 \begin{align}
\phi(r) =& i \pi\Bigg[-\frac{\alpha C_4+m^2C_1  }{(d-1) r_0^2}+\frac{m^2c_0 l^{d-2} \left(c_0 c_2
 	(d-4) (d-2)^2+c_1 (d-3) r_0\right)}{(d-3) (d-2)^2 (d-1)r_0^2 \Lambda}\Bigg]\log (r-r_0)+ ...
\end{align} 
 ... means finite terms.\\
Regularity on the horizon condition tells us to eliminate $\log(r-r_0)$. 
It causes $C_1$ to be as follows,\\
\begin{align} \label{C_1}
 C_1\to \frac{c_0 d_1 l^{d_4}  \left(c_0 c_2 d_4 d_2^2+c_1 d_3 r_0\right)}{2d_3 d_2^2m^2\Lambda}-\frac{\alpha
 	C_4}{m^2}
\end{align}    
By inserting Eq.(\ref{C_1}) into Eq.(\ref{Phi-}), $\phi(r)$ is as,
\begin{align}
\phi(r)& =\Phi_0+m^2 \Bigg(\frac{c_0  l^{d_2}}{2 d_3 d_2^2 d_1 \Lambda  r_0^4} \left(2 \Lambda  l^2 r_0^d \left(d \log r-\log \left(r r_0^d-r_0 r^d\right)\right) \left(c_0
	c_2 d_4 d_2^2+c_1 d_3 r_0\right)\right)\nonumber\\&+\frac{c_0 l^{d_2}}{2 d_3 d_2^2 d_1 \Lambda  r_0^4}d_2 d_1^2 r_0^5 \int \frac{r^{d_3} \left(c_0 c_2 d_1
		d_2^2+c_1 d_3 r\right)}{r r_0^d-r_0 r^d} \, dr\Bigg)=\Phi_0-m^2B(r)
\end{align}  
$\Phi_0$ is determined by applying $\phi(r \to \infty)=1$,
\begin{equation}\label{Phi_0}
\Phi_0= \phi(r \to \infty)+m^2B(r \to \infty)=1+m^2B(r \to \infty)
\end{equation}
By inserting $\Phi_0$ into Eq.((\ref{Phi-}),
we will have,
\begin{equation}
 \phi(r)=1+m^2B(r \to \infty)-m^2B(r)
\end{equation}
So the mode equation Eq.(\ref{EoM2}) is solved up to the first order in terms of $\alpha$ and $m^2$. Now we calculate the value of $\frac{\eta}{s}$ by using Eq.(\ref{formula}), 
 \begin{align}\label{result}
 \frac{\eta}{s}&= \frac{1}{4\pi}\phi(r_0)^2 = \frac{1}{4\pi} \Bigg(1+2m^2\bigg(B(r \to \infty)-B(r_0)\bigg)+O(m^4) \Bigg)
 \end{align}
It shows massive gravity with quintessence behaves like pure AdS massive gravity\cite{Hartnoll:2016tri}. Thus KSS bound is violated for this black brane  by applying $\phi(r_0)=\text{finite}$ and $\phi(r \to \infty)=1$.
 \section{Conclusion}
\noindent We studied the dual of $d$-dimensional AdS black brane solution surrounded by quintessence in massive gravity. Hydrodynamics is an effective theory of feild theory in the long wavelength limit. we calculate $\frac{\eta}{s}$ for this gravitational model as an important transport coefficients by applying the Dirichlet boundary and regularity on the horizon conditions\cite{Hartnoll:2016tri}. There is a conjecture that states $\frac{\eta }{s}=\frac{1}{4\pi}$ for Einstein-Hilbert gravity, known as KSS bound \cite{Kovtun2004} which it is violated for higher derivative gravity\cite{Ref23,Ref24,Sadeghi:2015vaa,Parvizi:2017boc,Sadeghi:2018vrf}. Our result shows that it's independent of quintessence in arbitrary dimensions. Since $\frac{\eta}{s}$ is
proportional to the inverse squared of the coupling in feild theory side. Our outcome also shows that quintessence acts like Yang-Mills charge \cite{Sadeghi:2018ylh} and cloud of string \cite{Sadeghi:2019muh}.\\\\
\noindent {\large {\bf Acknowledgment} }  Author would like to thank Shahrokh Parvizi and Komeil Babaei for useful comments and suggestions.



\begin{thebibliography}{}


\bibitem{deRham:2010kj} 
C.~de Rham, G.~Gabadadze and A.~J.~Tolley,
``Resummation of Massive Gravity,''
Phys.\ Rev.\ Lett.\  {\bf 106}, 231101 (2011)
[arXiv:1011.1232 [hep-th]].



\bibitem{Perlmutter:1998np} 
  S.~Perlmutter {\it et al.} [Supernova Cosmology Project Collaboration],
  ``Measurements of $\Omega$ and $\Lambda$ from 42 high redshift supernovae,''
  Astrophys.\ J.\  {\bf 517}, 565 (1999)
  doi:10.1086/307221
  [astro-ph/9812133].


\bibitem{Riess:1998cb} 
  A.~G.~Riess {\it et al.} [Supernova Search Team],
  ``Observational evidence from supernovae for an accelerating universe and a cosmological constant,''
  Astron.\ J.\  {\bf 116}, 1009 (1998)
  doi:10.1086/300499
  [astro-ph/9805201].
      
    \bibitem{Caldwell:1997ii} 
      R.~R.~Caldwell, R.~Dave and P.~J.~Steinhardt,
      Phys.\ Rev.\ Lett.\  {\bf 80}, 1582 (1998)
      doi:10.1103/PhysRevLett.80.1582
      [astro-ph/9708069].

 \bibitem{Kiselev:2002dx} 
   V.~V.~Kiselev,
   ``Quintessence and black holes,''
   Class.\ Quant.\ Grav.\  {\bf 20}, 1187 (2003)
   doi:10.1088/0264-9381/20/6/310
   [gr-qc/0210040].


\bibitem{Chen:2008ra} 
  S.~Chen, B.~Wang and R.~Su,
  ``Hawking radiation in a $d$-dimensional static spherically-symmetric black Hole surrounded by quintessence,''
  Phys.\ Rev.\ D {\bf 77}, 124011 (2008)
  doi:10.1103/PhysRevD.77.124011
  [arXiv:0801.2053 [gr-qc]].


\bibitem{Azreg-Ainou:2014lua} 
  M.~Azreg-Aïnou,
``Charged de Sitter-like black holes: quintessence-dependent enthalpy and new extreme solutions,''
  Eur.\ Phys.\ J.\ C {\bf 75}, no. 1, 34 (2015)
  doi:10.1140/epjc/s10052-015-3258-3
  [arXiv:1410.1737 [gr-qc]].

\bibitem{Ghosh:2016ddh} 
  S.~G.~Ghosh, M.~Amir and S.~D.~Maharaj,
  ``Quintessence background for 5D Einstein-Gauss-Bonnet black holes,''
  Eur.\ Phys.\ J.\ C {\bf 77}, no. 8, 530 (2017)
  doi:10.1140/epjc/s10052-017-5099-8
  [arXiv:1611.02936 [gr-qc]].

\bibitem{Ghosh:2017cuq} 
  S.~G.~Ghosh, S.~D.~Maharaj, D.~Baboolal and T.~H.~Lee,
  ``Lovelock black holes surrounded by quintessence,''
  Eur.\ Phys.\ J.\ C {\bf 78}, no. 2, 90 (2018)
  doi:10.1140/epjc/s10052-018-5570-1
  [arXiv:1708.03884 [gr-qc]].

\bibitem{Chabab:2020ejk} 
  M.~Chabab and S.~Iraoui,
  ``Thermodynamic criticality of d-dimensional charged AdS black holes surrounded by quintessence with a cloud of strings background,''
  arXiv:2001.06063 [hep-th].

  
\bibitem{Maldacena97}
J. M. Maldacena, ``The Large N limit of superconformal field theories and supergravity,'' Int.\ J.\ Theor.\ Phys.\  {\bf 38} (1999) 1113 [Adv.\ Theor.\ Math.\ Phys.\  {\bf 2} (1998) 231] [hep-th/9711200].

\bibitem{Bhattacharyya07}
S.~Bhattacharyya, V.~E.~Hubeny, S.~Minwalla and M.~Rangamani,
``Nonlinear Fluid Dynamics from Gravity,''
JHEP {\bf 0802}, 045 (2008)
[arXiv:0712.2456 [hep-th]].


\bibitem{Bhattacharya2011}
J.~Bhattacharya, S.~Bhattacharyya, S.~Minwalla and A.~Yarom,
``A Theory of first order dissipative superfluid dynamics,''
JHEP {\bf 1405}, 147 (2014)
[arXiv:1105.3733 [hep-th]]

\bibitem{Shuryak:2008eq} 
  E.~Shuryak,
  ``Physics of Strongly coupled Quark-Gluon Plasma,''
  Prog.\ Part.\ Nucl.\ Phys.\  {\bf 62}, 48 (2009)
  doi:10.1016/j.ppnp.2008.09.001
  [arXiv:0807.3033 [hep-ph]].

\bibitem{Cao:2010wa} 
  C.~Cao, E.~Elliott, J.~Joseph, H.~Wu, J.~Petricka, T.~Schäfer and J.~E.~Thomas,
 ``Universal Quantum Viscosity in a Unitary Fermi Gas,''
  Science {\bf 331}, 58 (2011)
  doi:10.1126/science.1195219
  [arXiv:1007.2625 [cond-mat.quant-gas]].

\bibitem{Policastro2001}
G.~Policastro, D.~T.~Son and A.~O.~Starinets,
``The Shear viscosity of strongly coupled $ \mathcal{N}=4 $ supersymmetric Yang-Mills plasma,''
Phys.\ Rev.\ Lett.\  {\bf 87}, 081601 (2001)
[hep-th/0104066].


\bibitem{Policastro2002}
G.~Policastro, D.~T.~Son and A.~O.~Starinets,``From AdS/CFT correspondence to hydrodynamics,''
JHEP {\bf 0209}, 043 (2002)
[hep-th/0205052]. 


\bibitem{Son2007}
D.~T.~Son and A.~O.~Starinets,
``Viscosity, Black Holes, and Quantum Field Theory,''
Ann.\ Rev.\ Nucl.\ Part.\ Sci.\  {\bf 57}, 95 (2007)
[arXiv:0704.0240 [hep-th]].



\bibitem{Kovtun2004}
P.~Kovtun, D.~T.~Son and A.~O.~Starinets,
``Viscosity in strongly interacting quantum field theories from black hole physics,''
Phys.\ Rev.\ Lett.\  {\bf 94}, 111601 (2005)
[hep-th/0405231].

\bibitem{Kovtun2012}
P. Kovtun,``Lectures on hydrodynamic fluctuations in relativistic theories,''J.\ Phys.\ A {\bf 45} (2012) 473001[arXiv:1205.5040 [hep-th]].

\bibitem{Ref23}
M.~Brigante, H.~Liu, R.~C.~Myers, S.~Shenker and S.~Yaida,
   ``The Viscosity Bound and Causality Violation,''
   Phys.\ Rev.\ Lett.\  {\bf 100}, 191601 (2008)
   [arXiv:0802.3318 [hep-th]].

 \bibitem{Ref24}
  I.~P.~Neupane and N.~Dadhich,
     ``Entropy Bound and Causality Violation in Higher Curvature Gravity,''
     Class.\ Quant.\ Grav.\  {\bf 26}, 015013 (2009)
     [arXiv:0808.1919 [hep-th]].

 

\bibitem{Sadeghi:2015vaa} 
  M.~Sadeghi and S.~Parvizi,
  ``Hydrodynamics of a black brane in Gauss-Bonnet massive gravity,''
  Class.\ Quant.\ Grav.\  {\bf 33}, no. 3, 035005 (2016)
  [arXiv:1507.07183 [hep-th]].


\bibitem{Parvizi:2017boc} 
S.~Parvizi and M.~Sadeghi,
``Holographic Aspects of a Higher Curvature Massive Gravity,''
Eur.\ Phys.\ J.\ C {\bf 79}, no. 2, 113 (2019)
doi:10.1140/epjc/s10052-019-6631-9
[arXiv:1704.00441 [hep-th]].

\bibitem{Sadeghi:2018vrf} 
M.~Sadeghi,
``Black Brane Solution in Rastall AdS Massive Gravity and Viscosity Bound,''
Mod.\ Phys.\ Lett.\ A {\bf 33}, no. 37, 1850220 (2018)
doi:10.1142/S0217732318502206
[arXiv:1809.08698 [hep-th]].

\bibitem{Burikham:2016roo} 
  P.~Burikham and N.~Poovuttikul,
  ``Shear viscosity in holography and effective theory of transport without translational symmetry,''
  Phys.\ Rev.\ D {\bf 94}, no. 10, 106001 (2016)
  doi:10.1103/PhysRevD.94.106001
  [arXiv:1601.04624 [hep-th]].


\bibitem{Cai:2014znn} 
R.~G.~Cai, Y.~P.~Hu, Q.~Y.~Pan and Y.~L.~Zhang,
``Thermodynamics of Black Holes in Massive Gravity,''Phys.\ Rev.\ D {\bf 91}, no. 2, 024032 (2015)
  doi:10.1103/PhysRevD.91.024032
[arXiv:1409.2369 [hep-th]]

  


\bibitem{Kovtun2003}
P.~Kovtun, D.~T.~Son and A.~O.~Starinets,
``Holography and hydrodynamics: Diffusion on stretched horizons,''
JHEP {\bf 0310}, 064 (2003)
[hep-th/0309213].

\bibitem{Hartnoll:2016tri} 
S.~A.~Hartnoll, D.~M.~Ramirez and J.~E.~Santos,
``Entropy production, viscosity bounds and bumpy black holes,''
JHEP {\bf 1603}, 170 (2016)
doi:10.1007/JHEP03(2016)170
[arXiv:1601.02757 [hep-th]].

\bibitem{Sadeghi:2018ylh} 
M.~Sadeghi,
``Einstein-Yang-Mills AdS black brane solution in massive gravity and viscosity bound,''
Eur.\ Phys.\ J.\ C {\bf 78}, no. 10, 875 (2018)
doi:10.1140/epjc/s10052-018-6360-5
[arXiv:1810.09242 [hep-th]].


  \bibitem{Sadeghi:2019muh} 
  M.~Sadeghi and H.~Ranjbari,
  ``Does a Cloud of Strings Affect Shear Viscosity Bound?,''
  Class.\ Quant.\ Grav.\  {\bf 36}, no. 20, 205012 (2019)
  doi:10.1088/1361-6382/ab436e
  [arXiv:1905.12856 [hep-th]].
  






















      
        













 

\end{thebibliography}
\end{document}